\documentclass{physeauth}
\usepackage{graphicx}
\usepackage{amsmath}
\usepackage{amssymb}

\begin{document}

\begin{frontmatter}

\title{Event-by-event simulation of quantum phenomena}

\author[rug]{H. De Raedt\thanksref{thank1}},
\author[rug]{S. Zhao},
\author[rug]{S. Yuan},
\author[rug]{F. Jin},
\author[embd]{K. Michielsen},
\author[tokyo]{S. Miyashita}

\address[rug]{Department of Applied Physics, Zernike Institute for Advanced Materials,
University of Groningen, Nijenborgh 4, 9747 AG Groningen, The Netherlands}
\address[embd]{EMBD, Vlasakker 21, 2160 Wommelgem, Belgium}
\address[tokyo]{Department of Physics, Graduate School of Science,
The University of Tokyo, 7-3-1 Hongo, Bunkyo-Ku, Tokyo 113-8656, Japan}

\thanks[thank1]{Corresponding author.
E-mail: h.a.de.raedt@rug.nl}

\begin{abstract}
We discuss recent progress in the development of simulation algorithms that
do not rely on any concept of quantum theory but are nevertheless capable of reproducing
the averages computed from quantum theory through an event-by-event simulation.
The simulation approach is illustrated by applications to 
Einstein-Podolsky-Rosen-Bohm experiments with photons.
\end{abstract}


\end{frontmatter}

\def\sumprime{\mathop{{\sum}'}}
\def\DLM{DLM}
\def\DLMS{DLMs}
\def\Eq#1{(\ref{#1})}

\section{Introduction}\label{Introduction}

Computer simulation is widely regarded as complementary to theory and experiment~\cite{LAND00}.
The standard approach is to start from one or more basic equations of physics and
to employ a numerical algorithm to solve these equations.
This approach has been highly successful for a wide variety of problems
in science and engineering.
However, there are a number of physics problems, very fundamental ones,
for which this approach fails, simply because there are no basic
equations to start from.

\begin{figure*}[t]
\begin{center}
\includegraphics[width=15cm]{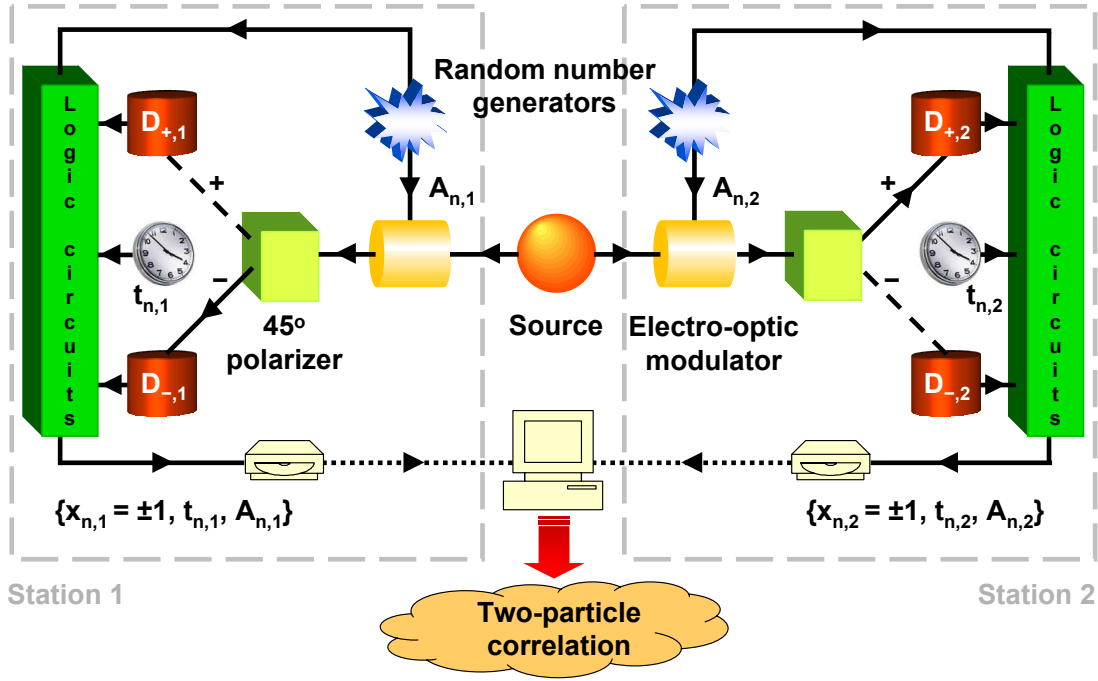}
\caption{(color online) Schematic diagram of an EPRB experiment with photons.
}
\label{fig1a}
\end{center}
\end{figure*}

Indeed, as is well-known from the early days in the development of quantum theory,
quantum theory has nothing to say about individual events~\cite{BOHM51,HOME97,BALL03}.
Reconciling the mathematical formalism that does not describe individual events
with the experimental fact that each observation yields a definite outcome
is referred to as the quantum measurement paradox and is
the most fundamental problem in the foundation of quantum theory~\cite{HOME97}.

In view of the quantum measurement paradox, it is unlikely
that we can find algorithms that simulate the experimental observation
of individual events within the framework of quantum theory.
Of course, we could simply use pseudo-random numbers to generate events according to the probability distribution
that is obtained by solving the time-independent Schr{\"o}dinger equation.
However, the challenge is to find algorithms that simulate, event-by-event,
the experimental observations of, for instance, interference without
first solving the Schr{\"o}dinger equation.

This paper is not about a new interpretation or an extension of quantum theory.
The proof that there exist simulation algorithms that reproduce the results of quantum theory
has no direct implications on the foundations of quantum theory: These algorithms describe the process of generating events
on a level of detail about which quantum theory has nothing to say~\cite{HOME97,BALL03}.
The average properties of the data may be in perfect agreement with quantum theory but
the algorithms that generate such data are outside of the scope of what quantum theory can describe.

In a number of recent papers~\cite{RAED05b,RAED05c,RAED05d,MICH05,RAED06c,RAED07a,RAED07b,RAED07c,RAED07d,ZHAO07b,ZHAO08,ZHAO08b,JIN08},
we have demonstrated that locally-connected networks of processing units
can simulate event-by-event, the single-photon beam splitter and Mach-Zehnder interferometer
experiments, universal quantum computation,
real Einstein-Podolsky-Rosen-Bohm (EPRB) experiments,
Wheeler's delayed choice experiment and the double-slit experiment with photons.
Our work suggests that we may have discovered a procedure to simulate quantum phenomena
using event-based, particle-only processes that satisfy Einstein's criterion of local causality,
without first solving a wave equation.
In this paper, we limit the discussion to event-by-event simulations of real EPRB experiments.

\section{EPRB experiments}
\label{EPRBexperiment}

In Fig.~\ref{fig1a}, we show a schematic diagram of an EPRB experiment
with photons (see also Fig.~2 in~\cite{WEIH98}).
The source emits pairs of photons.
Each photon of a pair propagates to an observation station
in which it is manipulated and detected.
The two stations are separated spatially and temporally~\cite{WEIH98}.
This arrangement prevents the observation at
station 1 (2) to have a causal effect on the
data registered at station $2$ (1)~\cite{WEIH98}.
As the photon arrives at station $i=1,2$, it passes through an electro-optic
modulator that rotates the polarization of the photon by an angle depending
on the voltage applied to the modulator.
These voltages are controlled by two independent binary random number generators.
As the photon leaves the polarizer, it generates a signal in one of the
two detectors.
The station's clock assigns a time-tag to each generated signal.
Effectively, this procedure discretizes time in intervals of a width that is
determined by the time-tag resolution $\tau$~\cite{WEIH98}.
In the experiment, the firing of a detector is regarded as an event.

As we wish to demonstrate
that it is possible to reproduce the results of quantum theory (which implicitly assumes idealized conditions)
for the EPRB gedanken experiment by an event-based simulation algorithm,
it would be logically inconsistent to ``recover'' the results of the former
by simulating nonideal experiments.
Therefore, we consider ideal experiments only,
meaning that we assume that detectors operate with 100\% efficiency,
clocks remain synchronized forever, the ``fair sampling'' assumption is satisfied~\cite{ADEN07}, and so on.
We assume that the two stations are separated spatially and temporally
such that the manipulation and observation at station 1 (2) cannot have a causal effect on the
data registered at station $2$ (1).
Furthermore, to realize the EPRB gedanken experiment on the computer,
we assume that the orientation of each electro-optic modulator
can be changed at will, at any time.
Although these conditions are very difficult to satisfy in real experiments,
they are trivially realized in computer experiments.

In the experiment, the firing of a detector is regarded as an event.
At the $n$th event, the data recorded on a hard disk at station $i=1,2$
consists of $x_{n,i}=\pm 1$, specifying which of the two detectors fired,
the time tag $t_{n,i}$ indicating the time at which a detector fired,
and the two-dimensional unit vector ${\bf a}_{n,i}$ that represents the rotation
of the polarization by the electro-optic modulator.
Hence, the set of data collected at station $i=1,2$ during a run of $N$ events
may be written as
\begin{eqnarray}
\label{Ups}
\Upsilon_i=\left\{ {x_{n,i} =\pm 1,t_{n,i},{\bf a}_{n,i} \vert n =1,\ldots ,N } \right\}
.
\end{eqnarray}
In the (computer) experiment, the data $\{\Upsilon_1,\Upsilon_2\}$ may be analyzed
long after the data has been collected~\cite{WEIH98}.
Coincidences are identified by comparing the time differences
$\{ t_{n,1}-t_{m,2} \vert n,m =1,\ldots ,N \}$ with a time window $W$~\cite{WEIH98}. 
Introducing the symbol $\sum'$ to indicate that the sum
has to be taken over all events that satisfy
$\mathbf{a}_i=\mathbf{a}_{n,i}$ for $i=1,2$,
for each pair of directions $\mathbf{a}_1$ and $\mathbf{a}_2$ of the electro-optic modulators,
the number of coincidences $C_{xy}\equiv C_{xy}(\mathbf{a}_1,\mathbf{a}_2)$ between detectors $D_{x,1}$ ($x =\pm 1$) at station
1 and detectors $D_{y,2}$ ($y =\pm1 $) at station 2 is given by
\begin{eqnarray}
\label{Cxy}
C_{xy}&=&\sumprime_{n,m=1}^N\delta_{x,x_{n ,1}} \delta_{y,x_{m ,2}}
\Theta(W-\vert t_{n,1} -t_{m ,2}\vert)
,
\end{eqnarray}
where $\Theta (t)$ is the Heaviside step function.
We emphasize that we count
all events that, according to the same criterion as the one employed in experiment,
correspond to the detection of pairs.
The average single-particle counts and the two-particle average are defined by
\begin{eqnarray}
E_1(\mathbf{a}_1,\mathbf{a}_2)&=&
\frac{\sum_{x,y=\pm1} xC_{xy}}{\sum_{x,y=\pm1} C_{xy}}
,\nonumber \\
E_2(\mathbf{a}_1,\mathbf{a}_2)&=&\frac{\sum_{x,y=\pm1} yC_{xy}}{\sum_{x,y=\pm1} C_{xy}}
,
\label{Ex}
\end{eqnarray}
and
\begin{eqnarray}
E(\mathbf{a}_1,\mathbf{a}_2)&=&
\frac{\sum_{x,y=\pm1} xyC_{xy}}{\sum_{x,y=\pm1} C_{xy}}
\nonumber \\
&=&\frac{C_{++}+C_{--}-C_{+-}-C_{-+}}{C_{++}+C_{--}+C_{+-}+C_{-+}}
,
\label{Exy}
\end{eqnarray}
respectively.
In Eqs.~(\ref{Ex}) and (\ref{Exy}), the denominator is the sum of all coincidences.

 For later use, it is expedient to introduce the function
 \begin{equation}
 \label{Sab}
 S(\mathbf{a},\mathbf{b},\mathbf{c},\mathbf{d})=
 E(\mathbf{a},\mathbf{c})-E(\mathbf{a},\mathbf{d})
 +
 E(\mathbf{b},\mathbf{c})+E(\mathbf{b},\mathbf{d})
 ,
 \end{equation}
 and its maximum
 \begin{eqnarray}
 \label{Smax}
 S_{max}&\equiv&\max_{\mathbf{a},\mathbf{b},\mathbf{c},\mathbf{d}} S(\mathbf{a},\mathbf{b},\mathbf{c},\mathbf{d})
 .
 \end{eqnarray}
 %
\subsection{Analysis of real experimental data}
\label{IIG}

We illustrate the procedure of data analysis and
the importance of the choice of the time window
$W$ by analyzing a data set (the archives
Alice.zip and Bob.zip) of an EPRB experiment with photons
that is publicly available~\cite{WEIHdownload}.

In the real experiment, the number of events detected at station 1 is unlikely
to be the same as the number of events detected at station 2.
In fact, the data sets of Ref.~\cite{WEIHdownload} show that
station 1 (Alice.zip) recorded 388455 events while
station 2 (Bob.zip) recorded  302271 events.
Furthermore, in the real EPRB experiment, there may be an
unknown shift $\Delta$ (assumed to be constant during the experiment)
between the times $t_{n,1}$ gathered at station 1 and
the times $t_{m,2}$ recorded at station 2.
Therefore, there is some extra ambiguity in
matching the data of station 1 to the data of station 2.

A simple data processing procedure that resolves this
ambiguity consists of two steps~\cite{WEIH00}.
First, we make a histogram of the time differences
$t_{n,1}-t_{m,2}$ with a small but reasonable resolution
(we used $0.5$ ns).
Then, we fix the value of the time-shift $\Delta$
by searching for the time difference for which
the histogram reaches its maximum, that is we maximize
the number of coincidences by a suitable choice of $\Delta$.
For the case at hand, we find $\Delta=4$ ns.
Finally, we compute the coincidences,
the two-particle average, and $S_{max}$ using
the expressions given earlier.
The average times between two detection events is $2.5$ ms and $3.3$ ms
for Alice and Bob, respectively.
The number of coincidences (with double counts removed) is
13975 and 2899 for ($\Delta=4$ ns, $W=2$ ns) and
($\Delta=0$ , $W=3$ ns) respectively.

\begin{figure}[t]
\begin{center}
\includegraphics[width=7.5cm]{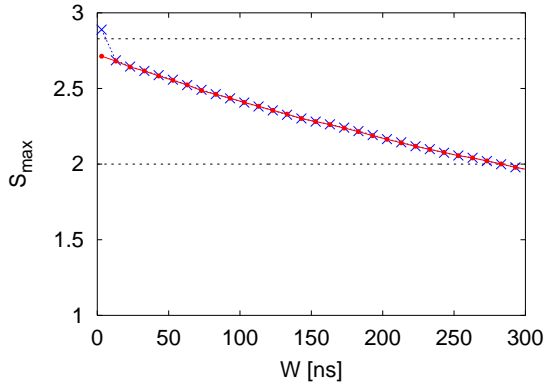}
\caption{(color online) $S_{max}$ as a function of the time window $W$,
computed from the data sets contained in the archives
Alice.zip and Bob.zip that can be downloaded from Ref.~\cite{WEIHdownload}.
Bullets (red): Data obtained by
using the relative time shift $\Delta=4$ ns that maximizes the
number of coincidences.
Crosses (blue): Raw data ($\Delta=0$).
Dashed line at $2\sqrt 2 $: $S_{max}$ if the system is described by quantum theory (see Section~\ref{Quantumtheory}).
Dashed line at $2$: $S_{max}$ if the system is described by the
class of models introduced by Bell~\cite{BELL93}.
}
\label{exp1}
\end{center}
\end{figure}

\begin{figure}[t]
\begin{center}
\includegraphics[width=7.5cm]{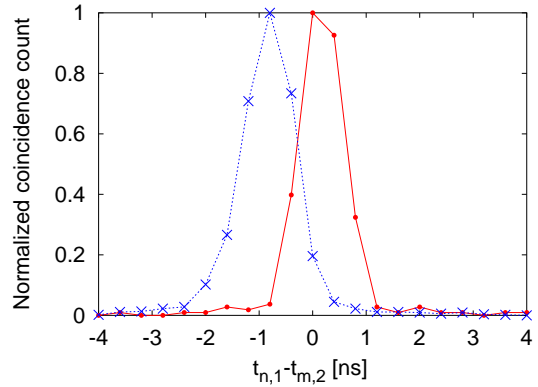}
\caption{(color online) Normalized coincidence counts as a function of time tag difference $t_{n,1}-t_{m,2}$,
computed from the data sets contained in the archives
Alice.zip and Bob.zip~\cite{WEIHdownload}, using the relative time shift $\Delta=4$ ns that maximizes the
number of coincidences.
Bullets (red): $\theta_1=0$ and $\theta_2=\pi/8$;
Crosses (blue): $\theta_1=0$ and $\theta_2=3\pi/8$.
}
\label{fort.7}
\end{center}
\end{figure}

In Fig.~\ref{exp1}  we present the results
for $S_{max}$ as a function of the time window $W$.
First, it is clear that $S_{max}$ decreases significantly
as $W$ increases but it is also clear that as $W\rightarrow0$,
$S_{max}$ is not very sensitive to the choice of $W$~\cite{WEIH00}.
Second, the procedure of maximizing the coincidence count
by varying $\Delta$ reduces the maximum value of $S_{max}$ from
a value 2.89 that considerably exceeds the maximum
for the quantum system ($2\sqrt{2}$, see Section~\ref{Quantumtheory}) to a value 2.73 that
violates the Bell inequality ($S_{max}\le2$, see Ref.~\cite{BELL93}) and is less than
the maximum for the quantum system.

Finally, we use the experimental data to show that the time delays depend on the
orientation of the polarizer. To this end, we
select all coincidences between $D_{+,1}$ and $D_{+,2}$ (see Fig.~\ref{fig1a}) and
make a histogram of the coincidence counts as a function of the time-tag difference,
for fixed orientation $\theta_1=0$ and the two orientations $\theta_2=\pi/8,3\pi/8$ (other
combinations give similar results).
The results of this analysis are shown in Fig.~\ref{fort.7}.
The maximum of the distribution shifts by approximately 1 ns as the polarizer at station 2 is
rotated by $\pi/4$, a demonstration that the time-tag data is sensitive to the orientation
of the polarizer at station 2. A similar distribution of time-delays (of about the same width) was also observed in a much older
experimental realization of the EPRB experiment~\cite{KOCH67}.
The time delays that result from differences in the orientations of the polarizers is much larger than
the average time between detection events, which for the data that we analyzed is about 30000 ns.
In other words, the loss in correlation that we observe as a function of increasing $W$ (see Fig.~\ref{exp1}) 
cannot be explained by assuming that we calculate correlations using photons that belong
to different pairs.

Strictly speaking, we cannot derive the time delay from classical electrodynamics:
The concept of a photon has no place in Maxwell's theory.
A more detailed understanding of the time delay mechanism
requires dedicated, single-photon retardation measurements for these specific optical elements.

\subsection{Role of the coincidence window $W$}

The crucial point is that in any real EPR-type experiment,
it is necessary to have an operational procedure to decide
if the two detection events correspond to the observation of
one two-particle system or to the observation of two single-particle systems.
In standard ``hidden variable'' treatments of the EPR gedanken experiment~\cite{BELL93},
the operational definition of ``observation of a single two-particle system'' is missing.
In EPRB-type experiments, this decision is taken on the basis of coincidence in time~\cite{KOCH67,CLAU74,WEIH98}.

Our analysis of the experimental data shows beyond doubt
that a model which aims to describe real EPRB experiments
should include the time window $W$ and that the interesting regime
is $W\rightarrow0$, not $W\rightarrow\infty$ as is assumed
in all textbook treatments of the EPRB experiment.
Indeed, in quantum mechanics textbooks it is standard to assume that an EPRB experiment
measures the correlation~\cite{BELL93}
\begin{eqnarray}
\label{CxyBell}
C_{xy}^{(\infty)}&=&\sumprime_{n=1}^N\delta_{x,x_{n ,1}} \delta_{y,x_{n ,2}}
,
\end{eqnarray}
which we obtain from Eq.~(\ref{Cxy}) by taking the limit $W\rightarrow\infty$.
Although this limit defines a valid theoretical model, there is no reason
why this model should have any bearing on the real experiments, in particular
because experiments pay considerable attention to the choice of $W$.
In experiments a lot of effort is made to reduce (not increase) $W$~\cite{WEIH98,WEIH00}.

\section{Quantum theory}
\label{Quantumtheory}
According to the axioms of quantum theory~\cite{BALL03},
repeated measurements on the two-spin system described by the density matrix $\rho$
yield statistical estimates
for the single-spin expectation values
\begin{eqnarray}
\label{Ei}
\widetilde E_1(\mathbf{a})&=&\langle \mathbf{\sigma}_1\cdot\mathbf{a} \rangle
\quad,\quad
\widetilde E_2(\mathbf{b})=\langle \mathbf{\sigma}_2\cdot\mathbf{b} \rangle
,
\end{eqnarray}
and the two-spin expectation value
\begin{eqnarray}
\label{Eab}
\widetilde E(\mathbf{a},\mathbf{b})&=&
\langle \mathbf{\sigma}_1\cdot\mathbf{a}\; \mathbf{\sigma}_2\cdot\mathbf{b} \rangle
,
\end{eqnarray}
where $\sigma_i=(\sigma_i^x ,\sigma_i^y ,\sigma_i^z )$
are the Pauli spin-1/2 matrices describing the spin of particle $i=1,2$~\cite{BALL03},
and $\mathbf{a}$ and $\mathbf{b}$ are unit vectors.
We have introduced the tilde to distinguish the quantum theoretical results from the results
obtained from the data sets $\{\Upsilon_1,\Upsilon_2\}$.

The quantum theoretical description of the EPRB experiment
assumes that the system is represented by the singlet state
$|\Psi\rangle=\left(| H\rangle_1 | V \rangle_2 -| V \rangle _1 |H \rangle_2\right)/\sqrt 2 $
of two spin-1/2 particles,
where $H$ and $V$ denote the horizontal
and vertical polarization and the subscripts refer to photon 1 and 2, respectively.
For the singlet state $\rho=|\Psi\rangle\langle\Psi|$,
$\widetilde E_1(\alpha)=\widetilde E_2(\beta)=0$ and
\begin{eqnarray}
\widetilde E(\alpha,\beta)&=&-\cos 2(\alpha -\beta)
\label{eq2}
.
\end{eqnarray}
%


\section{Simulation model}
\label{SimulationModel}

A concrete simulation model of the EPRB experiment sketched in Fig.~\ref{fig1a} requires
a specification of the information carried by the particles,
of the algorithm that simulates the source and
the observation stations, and of the procedure to analyze the data.
In the following, we describe a slightly modified version
of the algorithm proposed in Ref.~\cite{RAED06c}, tailored
to the case of photon polarization.

{\bf Source and particles:}
The source emits particles that carry a 
vector ${\bf S}_{n,i}=(\cos (\xi_{n}+(i-1)\pi/2) ,\sin (\xi_{n}+(i-1)\pi/2)$,
representing the polarization of the photons
that travel to station $i=1$ and station $i=2$, respectively.
Note that ${\bf S}_{n,1}\cdot {\bf S}_{n,2}=0$, indicating that
the two particles have orthogonal polarizations.
The ``polarization state'' of a particle is completely characterized by $\xi _{n}$,
which is distributed uniformly over the whole interval $[0,2\pi[$.
For the purpose of mimicing the apparent unpredictability of the experimental data,
we use uniform random numbers.
However, from the description of the algorithm, it will be clear that
the use of random numbers is not essential.
Simple counters that sample
the intervals $[0,2\pi[$ in a systematic, but uniform, manner might be employed
as well.

{\bf Observation station:}
The electro-optic modulator in station $i$ rotates ${\bf S}_{n,i}$ by an angle $\gamma _{n ,i}$, that is
${\bf a}_{n,i}=(\cos\gamma _{n ,i},\sin\gamma _{n ,i})$.
The number $M$ of different rotation angles is
chosen prior to the data collection (in the experiment of Weihs \textit{et al.}, $M=2$~\cite{WEIH98}).
We use $2M$ random numbers to fill the arrays $(\alpha _1 ,...,\alpha _M)$ and
$(\beta _1 ,...,\beta _M)$.
During the measurement process we use two uniform random numbers $1\le m,m'\le M$
to select the rotation angles $\gamma _{n,1} =\alpha _m$ and $\gamma _{n,2} =\beta _{{m}'}$.
The electro-optic modulator then rotates ${\bf S}_{n ,i} =(\cos (\xi _{n} +(i-1)\pi/2),\sin (\xi _{n}+(i-1)\pi/2 )$
by $\gamma _{n,i}$, yielding ${\bf S}_{n,i}=(\cos (\xi _{n} -\gamma_{n,i}+(i-1)\pi/2 ),\sin (\xi _{n} -\gamma_{n,i}+(i-1)\pi/2 )$.

The polarizer at station $i$ projects the rotated vector
onto its $x$-axis: ${\bf S}_{n ,i} \cdot \hat {\bf x}_i =\cos (\xi _{n } -\gamma _{n ,i}+(i-1)\pi/2 )$,
where $\hat {\bf  x}_i $ denotes the unit vector along the $x$-axis of the polarizer.
For the polarizing beam splitter, we consider a simple model: If $\cos ^2(\xi _{n } -\gamma _{n ,i}+(i-1)\pi/2 )>1/2$
the particle causes $D_{+1,i}$ to fire, otherwise $D_{-1,i}$ fires.
Thus, the detection of the particles generates the data
$x_{n ,i} =\mathop{\hbox{sign}}(\cos 2(\xi _{n } -\gamma _{n ,i} +(i-1)\pi/2))$.

{\bf Time-tag model:}
To assign a time-tag to each event,
we assume that as a particle passes through the detection system, it may experience a time delay.
In our model, the time delay ${t}_{n ,i} $ for a particle
is assumed to be distributed uniformly over the interval $[t_{0}, t_{0}+T]$,
an assumption that is not in conflict with available data~\cite{WEIH00}.
In practice, we use uniform random numbers
to generate ${t}_{n ,i}$.
As in the case of the angles $\xi_{n}$, the random choice of ${t}_{n ,i}$
is merely convenient, not essential.
From Eq.~(\ref{Cxy}), it follows that only differences of time delays matter.
Hence, we may put $t_0=0$.
The time-tag for the event $n$ is then $t_{n,i}\in[0,T]$.

There are not many options to make a reasonable choice for $T$.
Assuming that the particle ``knows'' its own direction
and that of the polarizer only, we can construct one
number that depends on the relative angle: ${\bf S}_{n ,i} \cdot \hat {\bf x}_i$.
Thus, $T=T(\xi _{n } -\gamma _{n ,i} )$ depends on $\xi _{n } -\gamma _{n ,i} $ only.
Furthermore, consistency with classical electrodynamics requires that
functions that depend on the polarization have period $\pi$~\cite{BORN64}.
Thus, we must have $T(\xi _{n } -\gamma _{n ,i} +(i-1)\pi/2)=F( ({\bf S}_{n ,i} \cdot \hat {\bf x}_i)^2)$.
We already used $\cos 2(\xi _{n } -\gamma _{n ,i}+(i-1)\pi/2 )$ to determine whether the particle generates
a $+1$ or $-1$ signal.
By trial and error, we found that $T(\xi _{n } -\theta_1 )=T_0 F(|\sin 2(\xi _{n } -\theta_1)|)=T_0|\sin 2(\xi _{n } -\theta_1)|^d$
yields useful results~\cite{RAED06c,RAED07a,RAED07b,RAED07c,RAED07d}.
Here, $T_0 =\max_\theta T(\theta)$ is the maximum time delay
and defines the unit of time, used in the simulation and $d$ is a free parameter of the model.
In our numerical work, we set $T_0=1$.

{\bf Data analysis:}
For fixed $N$ and $M$, the algorithm generates the
data sets $\Upsilon_i$
just as experiment does~\cite{WEIH98}.
In order to count the coincidences, we choose a
time-tag resolution $0<\tau<T_0 $ and a coincidence window $\tau\le W$.
We set the correlation counts $C_{xy} (\alpha _m ,\beta _{m'} )$ to zero for all $x,y=\pm 1$
and $m,m'=1,...,M$.
We compute the discretized time tags $k_{n ,i} =\lceil t_{n ,i}/ \tau\rceil $
for all events in both data sets.
Here $\lceil{x}\rceil$ denotes the smallest integer that is larger or equal to $x$, that is
$\lceil{x}\rceil-1<x\le\lceil{x}\rceil$.
According to the procedure adopted in the experiment~\cite{WEIH98},
an entangled photon pair is observed if and only if
$\left| {k_{n,1} -k_{n,2} } \right|<k=\lceil{W/\tau}\rceil$.
Thus, if $\left| {k_{n,1} -k_{n,2} } \right|<k$,
we increment the count $C_{x_{n,1},x_{n,2}} (\alpha _m ,\beta _{m'} )$.

\begin{figure}[t]
\begin{center}
\mbox{
\includegraphics[width=7.5cm]{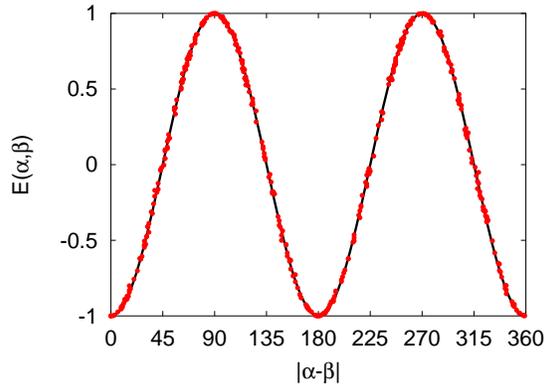}
}
\caption{(color online)
Comparison between computer simulation data (red bullets) and quantum theory
(black solid line) for the two-particle correlation $E(\alpha,\beta)$.
}
\label{fig2}
\label{fig4}
\end{center}
\end{figure}

\section{Simulation results}

The simulation proceeds in the same way as the experiment, that is we first
collect the data sets $\{\Upsilon_1,\Upsilon_2\}$, and then compute the coincidences
Eq.~(\ref{Cxy}) and the correlation Eq.~(\ref{Exy}).
The simulation results for the coincidences $C_{xy} (\alpha,\beta )$ depend on the
time-tag resolution $\tau$, the time window $W$ and the number of events $N$,
just as in real experiments~\cite{WEIH98}.

Figure~\ref{fig2} shows simulation data for $E(\alpha ,\beta )$
as obtained for d=2, $N=10^6$ and $W=\tau=0.00025T_0$.
In the simulation, for each event, the random numbers $1\le A_{n,i}\le M$ select one pair
out of $\{ (\alpha_i,\beta_j)|i,j=1,M\}$, where
the angles $\alpha_i$ and $\beta_j$ are fixed before the data is recorded.
The data shown has been obtained by allowing for $M=20$ different angles per station.
Hence, forty random numbers
from the interval [0,360[ were used to fill the arrays $(\alpha _1,\ldots,\alpha _M )$
and $(\beta _1 ,\ldots,\beta _M )$.
For each of the $N $ events, two different random number generators were used to select the
angles $\alpha _m $ and $\beta _{{m}'} $. The statistical correlation between $m$ and
$m'$ was measured to be less than $10^{-6}$.

From Fig.~\ref{fig2}, it is clear that the simulation data for $E(\alpha,\beta )$ are
in excellent agreement with quantum theory.
Within the statistical noise, the simulation data (not shown)
for the single-spin expectation values also
reproduce the results of quantum theory.

Additional simulation results (not shown) demonstrate that the kind of models described
earlier are capable of reproducing all the results of quantum theory for
a system of two S=1/2 particles~\cite{RAED06c,RAED07a,RAED07b,RAED07c,RAED07d}.
Furthermore, to first order in $W$ and in the limit that the number of events goes
to infinity, one can prove rigorously that these simulation models
give the same expressions for the single- and two-particle averages
as those obtained from quantum theory~\cite{RAED06c,RAED07a,RAED07b,RAED07c,RAED07d}.

\section{Discussion}

Starting from the factual observation that experimental realizations of the EPRB experiment
produce the data $\{\Upsilon_1,\Upsilon_2\}$ (see Eq.~(\ref{Ups})) and that coincidence in time is
a key ingredient for the data analysis,
we have described a computer simulation model that satisfies Einstein's criterion of local causality
and, exactly reproduces the correlation $\widetilde E(\mathbf{a}_1,\mathbf{a}_2)=-\mathbf{a}_1\cdot\mathbf{a}_2$
that is characteristic for a quantum system in the singlet state.

We have shown that whether or not these simulation models produce quantum correlations
depends on the data analysis procedure that is performed (long) after the data has been collected:
In order to observe the correlations of the singlet state,
the resolution $\tau$ of the devices that generate the
time-tags and the time window $W$ should be made as small as possible.
Disregarding the time-tag data ($d=0$ or $W>T_0$) yields results that disagree with quantum theory
but agree with the models considered by Bell~\cite{BELL93}.
Our analysis of real experimental data and our simulation
results show that increasing the time window changes the nature of the
two-particle correlations~\cite{RAED06c,RAED07a,RAED07b,RAED07c,RAED07d}.

According to the folklore about Bell's theorem, a procedure
such as the one that we described should not exist.
Bell's theorem states that any local, hidden variable
model will produce results that are in conflict with the quantum theory of a system of two $S=1/2$ particles~\cite{BELL93}.
However, it is often overlooked that this statement can be proven for a (very) restricted class of probabilistic models only.
In fact, Bell's theorem does not necessarily apply to the systems that we are interested in
as both simulation algorithms and actual data do not need to
satisfy the (hidden) conditions under which Bell's theorem hold~\cite{SICA99,HESS04,HESS05b,NIEU09}.

Furthermore, the apparent conflict between the fact that there exist event-based simulation models
that satisfy Einstein's criterion of local causality and reproduce
all the results of the quantum theory of a system of two $S=1/2$ particles
and the folklore about Bell's theorem, stating that such models are not supposed to exist
dissolves immediately if one recognizes that
Bell's extension of Einstein's concept of locality to the domain of probabilistic theories
relies on the hidden, fundamental assumption
that the absence of a causal influence implies logical independence~\cite{FINE74,JAYN89}.

The simulation model that is described in this paper
is an example of a purely ontological model that reproduces quantum phenomena
without first solving the quantum problem.
The salient features of our simulation models~\cite{RAED05b,RAED05c,RAED05d,MICH05,RAED06c,RAED07a,RAED07b,ZHAO07b} are that they
\begin{enumerate}
\item{generate, event-by-event, the same type of data as recorded in experiment,}
\item{analyze data according to the procedure used in experiment,}
\item{satisfy Einstein's criterion of local causality,}
\item{do not rely on any concept of quantum theory or probability theory,}
\item{reproduce the averages that we compute from quantum theory.}
\end{enumerate}

\section*{Acknowledgments}
We thank K. De Raedt and A. Keimpema for many useful suggestions and extensive discussions.


\end{document}